\documentclass[aip,apl,a4,preprint,twoside,floatfix,superscriptaddress]{revtex4-1} 
\usepackage{graphicx}
\usepackage{amsmath}
\usepackage{amssymb}
\usepackage{amsfonts}
\usepackage{dcolumn}
\usepackage{epsfig}
\usepackage{subfigure}
\usepackage{bm}
\usepackage[squaren,Gray]{SIunits}

\newcommand*{\Uchip}{\ensuremath{U_{\text{chip}}}}

\newcommand{\Rb}{\textsuperscript{87}Rb}
\newcommand{\ee}[1]{\ensuremath{\times 10^{#1}}}

\newcommand{\mm}{\milli\meter}
\newcommand{\nm}{\nano\meter}

\newcommand{\figwidth}{65mm}
\newcommand{\bigfigurewidth}{85mm}

\begin{document}





\title{Magneto-optical Trapping through a Transparent Silicon Carbide Atom Chip}








\newcommand{\tus}{Thales Underwater Systems, 525 route des Dolines, BP 157, Sophia-Antipolis, France}
\newcommand{\trt}{Thales Research and Technology France, Campus Polytechnique, 1 av. Fresnel, 91767 Palaiseau, France}
\newcommand{\lkb}{Laboratoire Kastler-Brossel, ENS, CNRS, Universit\'e Pierre et Marie Curie -- Paris 6, 24 rue Lhomond, 75005 Paris, France}
\newcommand{\IIIV}{III-V Lab, Route de Nozay, 91461 Marcoussis, France}

\author{Landry Huet}
\affiliation{\trt}
\affiliation{\tus}
\author{Mahdi Ammar}
\affiliation{\trt}
\affiliation{\lkb}
\author{Erwan Morvan}
\author{Nicolas Sarazin}
\affiliation{\IIIV}
\author{Jean-Paul Pocholle}
\affiliation{\trt}
\author{Jakob Reichel}
\affiliation{\lkb}
\author{Christine Guerlin}
\affiliation{\trt}
\affiliation{{Present address: LNE-SYRTE, Observatoire de Paris, 61 av. de l'Observatoire, 75014 Paris, France}}
\author{Sylvain Schwartz}
\affiliation{\trt}









\date{\today}

\begin{abstract}
We demonstrate the possibility of trapping about one hundred million rubidium atoms in a magneto-optical trap with several of the beams passing through a transparent atom chip mounted on a vacuum cell wall. The chip is made of a gold microcircuit deposited on a silicon carbide substrate, with favorable thermal conductivity. We show how a retro-reflected configuration can efficiently address the chip birefringence issues, allowing atom trapping at arbitrary distances from the chip. We also demonstrate detection through the chip, granting a large numerical aperture. This configuration is compared to other atom chip devices, and some possible applications are discussed.
\end{abstract}


\maketitle 

Atom chips\cite{Reichel1999,Folman2000} are a versatile tool for the manipulation of ultracold atoms.\cite{Fortagh2007} They have been successfully used to create atomic waveguides,\cite{Dekker2000} beam splitters\cite{Cassettari2000} or conveyor belts,\cite{Hansel2001a} and to achieve and handle Bose-Einstein condensates.\cite{Hansel2001b,Ott2001,Leanhardt2002} They opened new possibilities for the study of fundamental issues such as low-dimensional quantum systems,\cite{Hofferberth2007,Armijo2010} cavity quantum electrodynamics \cite{Colombe2007} and nanomechanical resonators.\cite{Hunger2010} Recent results regarding on-chip radiofrequency and microwave manipulation of atoms \cite{Schumm2005,Bohi2009,Deutsch2010} also hold prospects for future applications such as quantum information processing,\cite{Calarco2000,Treutlein2006} timekeeping\cite{Rosenbusch2009} or inertial sensing.\cite{Zatezalo2008}

One key feature of atom chips is to take advantage of near-field magnetic gradients, using the fact that atoms are trapped very close to the chip surface\cite{Fortagh2007} (typically on the order of tens or hundreds of microns). Because the chip size is usually centimetric, this comes at the price of reducing the optical access to the atoms to typically half the full $4\pi$ solid angle. In this paper, we address this issue by using a transparent atom chip made of a gold microcircuit deposited on a single crystal silicon carbide (SiC) substrate. In particular, we experimentally demonstrate trapping of atoms at arbitrary distances from the chip surface using a standard magneto-optical trap (MOT) configuration with several of the beams passing through the atom chip. We show how the polarization effects induced by the substrate birefringence can be robustly compensated. We finally illustrate the advantage of getting full optical access by detecting atomic fluorescence through the chip.

\begin{figure}[htb]
\includegraphics[width=\figwidth]{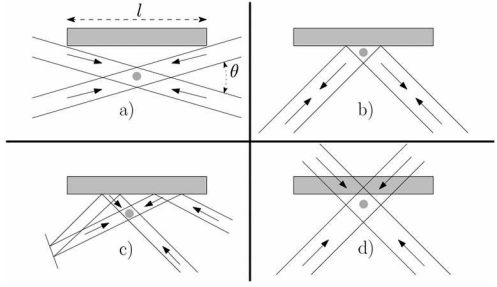} 
\caption{Different MOT configurations in the vicinity of an atom chip: angled MOT (a), mirror MOT (b), $\Lambda$ MOT (c) and MOT through a transparent chip (d). The chip is shown in grey and propagation directions of light are indicated by arrows.}
\label{FIG:ChipMOT}
\end{figure}

Before being loaded in a magnetic trap created by the atom chip, atoms are usually captured and cooled down using a MOT.\cite{Fortagh2007} The distance $d$ between the center of the MOT and the chip surface is an important parameter, since it determines the amount of electrical current required to transfer the atoms from the MOT to the chip magnetic trap, and the associated transfer efficiency.\cite{Squires2008} A first possibility for shortening $d$ in a classical 6-beam MOT configuration is to reduce the angle $\theta$ between two of the counter-propagating beam pairs. This so-called \emph{angled-MOT} is illustrated on figure \ref{FIG:ChipMOT}a. It has been shown\cite{Squires2008} however that $\theta$ cannot be reduced below a critical value $\theta_c \approx \unit{40}{\degree}$ without compromising the stability of the angled-MOT. This puts an absolute lower bound on the value of $d$, on the order of $(l/2)\tan(\theta_c/2)$, where $l$ is the width of the atom chip. For $l=\unit{15}{\mm}$, this leads to $d \gtrsim \unit{2.7}{\mm}.$ In practice, the MOT is even further from the chip due to the size of the beam waists. Consequently, the currents required to transfer the atoms to the chip magnetic trap are relatively strong.\cite{Squires2008} Another option, which results in a drastic reduction of $d$, is the so-called \emph{mirror-MOT},\cite{Reichel1999} as illustrated on figure \ref{FIG:ChipMOT}b. Such a configuration, where the atoms are trapped directly at the chip surface, requires planarization and reflective coating of the atom chip. In this case, the beam size imposes an upper bound on the value of $d,$ and the proximity of the chip limits the capture volume. Moreover, the polarization states of the reflected beams cannot be controlled independently, since they are linked by the reflection coefficients of the on-chip mirror. Some of the latter drawbacks can be circumvented in the so-called \emph{$\Lambda$ MOT},\cite{Ohadi2009} illustrated on figure \ref{FIG:ChipMOT}c.  In this configuration, an arbitrary value of $d$ can be achieved, as well as independent control over the polarization states. This comes however at the price of more complex opto-mechanics, and does not tackle the issues associated with planarization and reflective coatings. In the rest of this paper, we consider a fourth option where a classical MOT configuration is used with some of the beams propagating through a transparent atom chip, as illustrated on figure \ref{FIG:ChipMOT}d. This configuration should not be confused with the one described in Ref.~\onlinecite{Shevchenko2006}, which involves a permanent magnet transparent thin film.
In particular, the use of our electronic microcircuit allows both static and alternating (e.g. radiofrequency and microwave) magnetic fields. SiC is moreover rigid enough to be used as a part of the vacuum chamber itself.

Among all possible materials, single crystal SiC appears as a particularly relevant candidate substrate for such a configuration. With a bandgap value of about \unit{3.2}{\electronvolt} at room temperature,\cite{CREE} SiC is transparent at \unit{780}{\nm}. Its electrical resistivity (over \unit{10^5}{\ohm\usk\centi\meter} for high purity semi-insulating 4H SiC)\,\cite{CREE} and thermal conductivity (over \unit{390}{\watt\usk\reciprocal\meter\usk\reciprocal\kelvin} for 4H SiC)\,\cite{CREE} make SiC well suited for supporting wires with high electrical currents. In our experiment a $\unit{414}{\micro\meter} \times \unit{15}{\mm}\times\unit{15}{\mm}$ SiC atom chip is used. Gold wires are deposited on it following a pattern that has been proven elsewhere to achieve Bose-Einstein condensation of \Rb\ atoms.\cite{Farkas2010} The chip is coated with a dielectric anti-reflection (AR) layer on each side, allowing $80\%$ transmittance of the chip, including reflection by the wires.

\begin{figure}[htb]
\includegraphics[width=\figwidth]{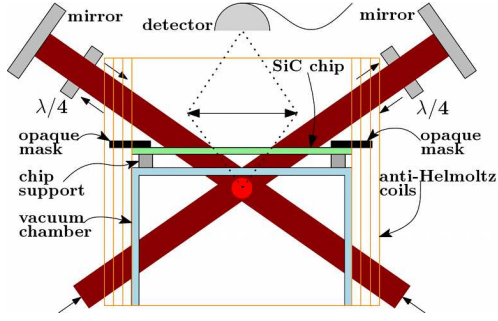} 
\caption{Sketch of the upper chamber part of the experimental setup. An additional pair of beams perpendicular to the plane of the figure is used to complete the 6-beam MOT configuration. Atomic fluorescence is recorded through the chip by the detection system (camera or photodiode). An opaque mask surrounding the chip ensures that only beams going through the chip contribute to the MOT.}
\label{FIG:ownSetup}
\end{figure}

The substrate we use has a 4H hexagonal crystalline structure and is cleaved with a $\unit{8}{\degree}10'$ deviation of the 1C axis. This grants the substrate uniaxial anisotropy, with ordinary optical index $n_o = 2.617$ and extraordinary optical index $n_e = 2.666$ at $\unit{780}{\nm}$.\cite{Shaffer1971} Hence special care is required on the MOT beams in order to get the correct circular polarizations on the trapped atoms.
Assuming no polarization-dependent losses (which can be made negligible by an appropriate AR coating), the effect of the atom chip on the beam polarization can be described by a Jones matrix of the form $\Uchip = R_{\alpha}G_{\gamma}R_{-\alpha},$ where $R_{\alpha}$ is a rotation matrix with angle $\alpha,$ and $G_{\gamma}=\left[\begin{matrix}e^{i\gamma/2} & 0\\ 0 & e^{-i\gamma/2}\end{matrix}\right].$ Using computations similar to those described in Ref.~\onlinecite{Vansteenkiste1993}, it can be proven that circular polarizations can be preserved in the retroreflection configuration sketched on Fig.~\ref{FIG:ownSetup} with only one single quarter-wave plate for each retro-reflected beam, provided the latter is oriented such that
\begin{align}
\phi &= \alpha - \frac{\pi}{4} , \label{formula:polar3}
\end{align}
where $\phi$ is the angle between the quarter-wave plate eigen basis and the reference polarization basis.
As can be seen in equation~\eqref{formula:polar3}, the optimal value of $\phi$ depends only on the chip polarization eigenbasis, and not on the phase shift $\gamma.$ One can therefore expect robustness to variations in the refractive indices and thickness of the chip, induced for instance by temperature changes, making it possible to precalibrate the quarter-wave plates outside the experimental setup. The quality of the AR coating is also very important to avoid intensity and polarization fluctuations linked to residual Fabry-Perot effects within the chip, leading to atom number fluctuations in the MOT. Our substrate showed the latter effect with a period of about \unit{11}{\celsius}, corresponding to $\frac{\partial n}{\partial T} \simeq \unit{9\ee{-5}}{\per\celsius}$ at \unit{780}{\nm} for 4H SiC between \unit{25}{\celsius} and \unit{65}{\celsius}, where $n=\left(n_o+n_e\right)/2$.

For the proof-of-concept experimental demonstration of the proposed architecture, the setup sketched on figure \ref{FIG:ownSetup} is used. The glass vacuum cell was manufactured by the company  ColdQuanta, with a differential vacuum system. A \Rb\  MOT is formed in the ultra-high vacuum part of the cell and is loaded from a 2D MOT with a push beam. Each laser beam contains a cooling beam and a repumping beam. The vapor pressure inside the cell is lower than \unit{10^{-11}}{\pascal}, as indicated by ion pump current measurements. Two coils in an anti-Helmoltz configuration are used. They create a \unit{0.24}{\tesla\usk\reciprocal\meter} magnetic field gradient at the field zero. The chip rests on a \unit{1}{\mm}-thick support on top of the glass cell. The two retroreflected beams cross the chip with a \unit{60}{\degree} incidence. Quarter-wave plates are placed between the chip and the retro-reflection mirrors, and are oriented according to equation \eqref{formula:polar3} to ensure the same circular polarization on the way back in the vacuum cell.

\begin{figure}[htb]
\includegraphics[width=\bigfigurewidth]{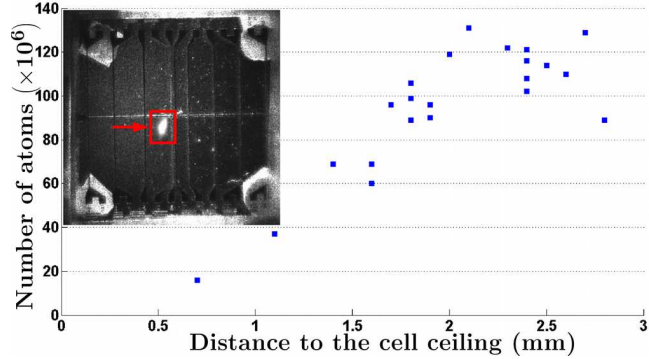} 
\caption{Main figure: variations of the number of atoms in the MOT with the distance of the cloud centroid to the cell ceiling. Inset: fluorescence imaging from the MOT atoms as seen through the chip. The picture was taken from above the vacuum cell using a Point Grey FL2-14S3M CCD camera.}
\label{Figure:combined}
\end{figure}

The possibility of imaging the MOT through the transparent chip is illustrated on the inset of Fig.~\ref{Figure:combined}. A convergent lens of \unit{50}{\mm} focal length and \unit{25.4}{\mm} diameter in a $2f-2f$ configuration with the atom cloud is used to improve the numerical aperture of the detection system. The number of atoms in the trap can be estimated from fluorescence measurements using a photodiode. Approximately \unit{1\ee{-6}}{\watt} of fluorescence is measured, over a solid angle of \unit{4\ee{-3}\times 4\pi}{\steradian}. Cooling light beams are detuned \unit{10.8}{\mega\hertz} away from resonance. All beams have an approximately \unit{13}{\mm} $1/e^2$ waist diameter. Atoms are illuminated with a total mean intensity on the order of \unit{80}{\milli\watt\per\centi\meter\squared}, as computed from the intensities measured at the center of the MOT beams and push beam, and from the transmission coefficients of the chip and of the cell glass walls. With a saturation intensity of \unit{4.1}{\milli\watt\per\centi\meter\squared} as given in Ref.~\onlinecite{Lewandowski2003}, this leads to an estimation\cite{Lewandowski2003} of $1.1\ee{8}$ atoms in the MOT. The characteristic loading and decay times of the trap are approximately \unit{170}{\milli\second} and \unit{460}{\milli\second} respectively.  The number of atoms is comparable to what was obtained in our previous setup, similar to the configuration presented here but without a chip. It also stands the comparison to the performances of other techniques of near-chip magneto-optical trapping.\cite{Squires2008,Reichel1999,Ohadi2009}

The atom number estimated above corresponds to a MOT center located \unit{1.5}{\mm} below the vacuum cell ceiling. By translating the quadrupole coils and realigning the optomechanical apparatus, we have studied the number of atoms as a function of the latter distance, as reported on Fig.~\ref{Figure:combined}. Positions of the cloud centroid are estimated by analyzing the images from a CCD camera on the cell side. As can be seen on Fig.~\ref{Figure:combined}, a consistent number of atoms is kept in the MOT at distances larger than \unit{1.5}{\mm}. This quickly drops when the cloud is closer to the cell ceiling, which is consistent with previous observations.\cite{Reichel1999} The typical size of our atom cloud is on the order of a few millimeters, thus the limitation of the cloud extension by the ceiling surface may account for the loss of atoms when the MOT is too close to the surface.

The possibility of trapping atoms near transparent chips could open the way to new applications combining for example the simplicity of chip evaporative cooling with more complex architectures requiring full optical access such as Ramsey-Bord\'e interferometers\cite{Bodart2010} or Bloch oscillators.\cite{Clade2005b} Detection through the chip could moreover be a powerful tool to improve numerical aperture for atom optical manipulation or in-situ detection with possible applications to on-chip atomic clocks or quantum information processing. In this respect, lenses etched directly on the SiC chip\cite{Lee2005} could combine a large numerical aperture with a particularly compact and scalable setup.

This work has been carried out within the CATS project ANR-09-NANO-039 funded by the French National Research Agency (ANR) in the frame of its 2009 program in Nanoscience, Nanotechnologies and Nanosystems (P3N2009).

\end{document}